\documentclass{elsart}

\usepackage{graphics}
\usepackage{graphicx}
\usepackage{epsfig}

\usepackage{amssymb}

\usepackage{slashbox}

\begin{document}

\begin{frontmatter}

\title{A General Construction  of   Binary  Sequences with Optimal Autocorrelation}
\thanks[label0]{This work is supported by the National Natural Science Foundations of China
(No.61170319), the Natural Science Fund of Shandong Province (No.ZR2010FM017), the Fundamental Research Funds for the Central Universities(No.11CX04056A) and the China Postdoctoral Science Foundation funded project (No.119103S148)}
\author[label1,label2]{Tongjiang Yan} \ead{yantoji@163.com }
\author[label2,label3]{Zhixiong Chen}
\author[label2]{Bao Li}
\address[label1]{College of Sciences, China University
of Petroleum, Qingdao 266555,  China.}
\address[label2]{State Key Laboratory of Information Security (Graduate
University of Chinese Academy of Sciences), Beijing 100049, China.}
\address[label3]{Department of Mathematics, Putian University, Putian, Fujian 351100, China}

\begin{abstract}
A general construction of binary sequences with low autocorrelation are considered in the paper. Based on recent progresses about this topic and this construction, several classes of binary sequences with optimal autocorrelation and other low autocorrelation are presented.
\end{abstract}


\end{frontmatter}
\newtheorem{theorem}{Theorem}
\newtheorem{lemma}{Lemma}
\newtheorem{corollary}{Corollary}
\newtheorem{property}{Property}
\newtheorem{definition}{Definition}
\newtheorem{proposition}{Proposition}
\newtheorem{remark}{Remark}
\section{Introduction}

$I_{0022}^{3}$
Pseudo-random sequences with low cross correlation can be employed in CDMA communications
to combat interference from the other users who
share a common channel and in stream cipher cryptosystems
as key stream generators to resist cross-correlation
attacks \cite{wg,brown}. Given two binary sequences $a=a(t)$ and $b=b(t)$ of period $N$, the periodic correlation between them is defined by
$$
R_{a, b}(\tau)=\sum\limits_{t=0}^{N-1}(-1)^{a(t)+b(t+\tau)}, 0\leq \tau < N,
$$
where the addition $t+\tau$ is performed modulo $N$. If $a=b$, $R_{a, b}(\tau)$
is called the (periodic) autocorrelation function of $a$, denoted by $R_{a}(\tau)$, or simply $R(\tau)$ if the context is clear, otherwise, $R_{a, b}(\tau)$
is called the (periodic) cross-correlation function of $a$ and $b$. Defining  the set
$$C_a=\{0\leq t\leq N-1:a(t)=1\}$$
the support of $a(t)$, then
\begin{eqnarray}\label{rakc}
R_{a}(\tau)=N-4(\mid C_a\mid-\mid (\tau+C_a)\cap C_a\mid).
\end{eqnarray}
The optimal values of out-of-phase autocorrelation of binary sequences in
terms of the smallest possible values of the autocorrelation are
classified into four types as follows: If $N\equiv 0\bmod 4$, $R(\tau)=\{0, -4, 4\}$; if $N\equiv 1\bmod 4$, $R(\tau)\in \{1, -3\}$;
if $N\equiv 2\bmod 4$, $R(\tau)\in \{2, -2\}$;
if $N\equiv 3\bmod 4$, $R(\tau)=-1$.
In the last case, $R(\tau)$ is often called ideal autocorrelation. For more details about optimal autocorrelation, the reader is referred  to \cite{kd,gg}.

In 1995, G. Gong found that most of the known sequences with ideal autocorrelation possess the following interleaved construction \cite{g}.
\begin{definition} Fix two positive integers $T$ and $K$ where $T \geq 2$ and $K \geq 1$. Given a binary sequence $a=(a(0), a(1), \ldots, a(K-1))$  of period $K$.  If the binary  sequence $u=(u(0), u(1), \ldots, u(KT-1))$ can be given by
an $K\times T$ matrix $A(u)$ as follows:
\begin{eqnarray} \label{uarray}
u=\left(
\begin{array}{ccccc}
u(0)&\ldots&u(T-1)\\
u(T)&\ldots&u(2T-1)\\
\vdots&&\vdots\\
u((K-1)T)&\ldots&u(KT-1)\\
\end{array}
\right)
\end{eqnarray}
which satisfies that each column of $A(u)$ is a shift of $a$ or a all zero sequence, then $u$ is called an interleaved sequence.
\end{definition}

Let $A_i$ be the $i$th column. Then $u=(A_0, \ldots, A_{T-1})$. With the development of interleaved technology, the above definition was generalized to the case that not all nonzero column vector $A_j$ are required to be shift equivalent. For example, in \cite{ng}, the case that $A_j$'s are equivalent to their complements can be permitted. For more details about the interleaved construction, the reader is referred to \cite{gg}. In the paper, we use the generalized definition of interleaved sequences. For the original  interleaved sequences, we call them  classical interleaved sequences.

Assume the binary sequence
$s(t)=I(a_{0}(k), a_1(k), a_2(k), \ldots, a_{T-1}(k))$
possess a $(K, T)$ interleaved construction, where each $a_{i}(k)$ is a binary column sequence of period $K$, and $L^{\tau}(s(t))$ denote the left $\tau-$ shift of $s(t)$ \cite{tg}. If $\tau=\tau_1T+\tau_2$, where $0\leq \tau_2\leq T-1$, then
\begin{lemma}\label{0} \cite{g} The array form of $L^{\tau}(s(t))$ is given by
\begin{eqnarray}\label{shift}
I(a_{\tau_2}(k+\tau_1),
\ldots, a_{T-1}(k+\tau_1),
a_{0}(k+\tau_1+1), \ldots, a_{\tau_2-1}(k+\tau_1+1)).
\end{eqnarray}
\end{lemma}

In 2001, K. T. Arasu, C. Ding, T. Helleseth, P. Kumar and H. Martinsen  gave a construction of binary sequences with  optimal autocorrelation of
period $4N$ by  sequences of
period $N\equiv 3\bmod 4$ with ideal autocorrelation \cite{kd}. Then this construction was generalized in \cite{zhang} and found to possess interleaved construction \cite{ng}. In 2010, X. Tang and G. Gong gave three new interleaved constructions of binary sequences with optimal autocorrelation values \cite{tg}. This paper will search more general constructions which can include them and some other new binary sequences with low autocorrelation.

\section{An Interleaved Sequence and Its Modification }

Define a pair of  binary sequences $s$ and  $s'$ by
\begin{itemize}
\item Construction A: $s=I(0_{K}, a_1, a_2, \ldots, a_{T-1}),$
\item Construction B: $s'=I(1_{K}, a_1, a_2, \ldots, a_{T-1})$,
\end{itemize}
where $0_{K}$ and $1_{K}$ are all zero sequence and all one sequence of period $K$ respectively,  $a_i'$s  are binary sequences of period $K$.The balance difference of  $a_i$  is given as $d(a_{i})=2\mid C_{a_{i}}\mid-K.$

In \cite{tg}, generalized GMW sequences and their modifications  of period $2^{2n}-1$ are defined respectively
as the above sequences $s$ and $s'$ with an additional condition that all $a_i'$s  are some shifts of ideal autocorrelation sequence $a$. Then  $d(a_i)$ is constant and takes value $1$ or $-1$. If $d(a_i)=-1$, we can get a pair of modified sequences $\bar{s}$ and $\bar{s'}$ by replacing each $a_i$ with its complement sequence, and keep their autocorrelation unchanged \cite{gg}. So we may assume that each $d(a_i)$ always takes the value $1$ when $s$ and $s'$ are generalized GMW sequences and their modifications respectively.

The sequence $s$ and its modification $s'$ have the following properties of correlation.

\begin{theorem}\label{ss} Let $\tau=\tau_1T+\tau_2,  0\leq \tau_2 \leq T-1$.
\begin{eqnarray*}
&&R_{s'}(\tau)=
\left\{
\begin{array}{lll}
R_{s}(\tau) & \mbox{ if } \tau_2=0,\\
R_{s}(\tau)+2d(a_{\tau_2})+2d(a_{T-\tau_2})&\mbox{ if } \tau_2\neq 0.
\end{array}
\right.
\end{eqnarray*}
The cross-correlation of $s$ and $s'$ is given by
\begin{eqnarray*}
&&R_{ss'}(\tau)=\left\{
\begin{array}{lll}
TK-2K & \mbox{ if } \tau=0,\\
R_{s}(\tau)-2K&\mbox{ if } \tau_2= 0, \tau\neq0, \\
R_{s}(\tau)+2d(a_{T-\tau_2})&\mbox{ otherwise };
\end{array}
\right.
\end{eqnarray*}

\begin{eqnarray*}
&&R_{s's}(\tau)=\left\{
\begin{array}{lll}
KT-2K & \mbox{ if } \tau=0,\\
R_{s}(\tau)-2K&\mbox{ if } \tau_2=0, \tau\neq0, \\
R_{s}(\tau)+2d(a_{\tau_2})&\mbox { otherwise }.
\end{array}
\right.
\end{eqnarray*}
\end{theorem}

\noindent\textbf{Proof. } To calculate $R_{s}(\tau)$, we need compare sequences $s=I(0_{K}, a_1, a_2, \ldots, a_{T-1})$ and its  $\tau-$ shift $L^{\tau}(s)$.

If $\tau_2=0$, from Lemma \ref{0},
$$L^{\tau}(s)=I(0_{K}, L^{\tau_1}(a_1),\ldots,  L^{\tau_1}(a_2), \ldots, L^{\tau_1}(a_{T-1})),$$
then \begin{eqnarray}\label{eq:rs}
R_s(\tau)=K+\sum\limits_{i=1}^{T-1}R_{a_i}(\tau_1).
\end{eqnarray}
Similarly,  we have $R_{s'}(\tau)=K+\sum\limits_{i=1}^{T-1}R_{a_i}(\tau_1)$. So  $R_s(\tau)=R_{s'}(\tau).$

If  $\tau_2\neq 0$, from Lemma \ref{0},
 $$L^{\tau}(s)=I(L^{\tau_1}(a_{\tau_2}),\ldots, L^{\tau_1}(a_{T-1}), 0_{K}, \ldots, L^{\tau_1+1}(a_{\tau_2-1})).$$
Thus we have
\begin{eqnarray}\label{eq:rst}
R_s(\tau)&=&\!\!\!\sum\limits_{i=1}^{T-\tau_2-1}R_{a_ia_{i+\tau_2}}(\tau_1)
+\!\!\!\sum\limits_{i=T-\tau_2+1}^{T-1}R_{a_ia_{i-(T-\tau_2)}}(\tau_1+1)\\
&&-d(a_{\tau_2})-d(a_{T-\tau_2}).\nonumber
\end{eqnarray}
Similarly
\begin{eqnarray}\label{eq:rsst}
R_{s'}(\tau)&=&\!\!\!\sum\limits_{i=1}^{T-\tau_2-1}R_{a_ia_{i+\tau_2}}(\tau_1) +\!\!\!\sum\limits_{i=T-\tau_2+1}^{T-1}R_{a_ia_{i-(T-\tau_2)}}(\tau_1+1)
\\
&&+d(a_{\tau_2})+d(a_{T-\tau_2}).\nonumber
\end{eqnarray}
From the above Equations (\ref{eq:rst}) and (\ref{eq:rsst}),  we have
$$R_{s'}(\tau)=R_{s}(\tau)+2d(a_{\tau_2})+2d(a_{T-\tau_2}).$$

To calculate  $R_{ss'}(\tau)$, we need compare sequences $s=I(0_{K}, a_1, a_2, \ldots, a_{T-1})$ and  $L^{\tau}(s')$, the $\tau-$ shift of $s'$.

If $\tau=0$, $s=I(0_{K}, a_1, a_2, \ldots, a_{T-1})$ compares with $s'=I(1_{K}, a_1, a_2, \ldots, a_{T-1})$, then
$$R_{ss'}(\tau)=\sum\limits_{k=0}^{K-1}(-1)^{0+1}+\sum\limits_{i=1}^{T-1}R_{a_i}(0)=-K+K(T-1)=KT-2K.$$

If $\tau_2=0, \tau\neq0$, $s=I(0_{K}, a_1, a_2, \ldots, a_{T-1})$ compares with $L^{\tau}(s')$, where
$$L^{\tau}(s')=I(1_{K}, L^{\tau_1}(a_1), L^{\tau_1}(a_2), \ldots, L^{\tau_1}(a_{T-1})),$$
then, from Equation (\ref{eq:rs}),
$$R_{ss'}(\tau)=\sum\limits_{k=0}^{K-1}(-1)^{0+1}+\sum\limits_{i=1}^{T-1}R_{a_i}(\tau_1)=R_{s}(\tau)-2K.$$

If $\tau_2\neq 0$, the $\tau-$ shift of $s'$

$$L^{\tau}(s')=I(L^{\tau_1}(a_{\tau_2}), L^{\tau_1}(a_{\tau_2+1}),\ldots,  L^{\tau_1}(a_{T-1}), 1_{K}, L^{\tau_1+1}(a_{1}), \ldots, L^{\tau_1+1}(a_{\tau_2-1})).$$

Then, from the comparison of $s$ and $L^{\tau}(s')$ and Equation (\ref{eq:rst}),
\begin{eqnarray}\label{eq:rssst}
R_{ss'}(\tau)&=&\sum\limits_{i=1}^{T-\tau_2-1}R_{a_ia_{i+\tau_2}}(\tau_1)
+\!\!\!\sum\limits_{i=T-\tau_2+1}^{T-1}R_{a_ia_{i-(T-\tau_2)}}(\tau_1+1)\\
&&-d(a_{\tau_2})+d(a_{T-\tau_2}).\nonumber\\
&=&R_{s}(\tau)+2d(a_{T-\tau_2}).\nonumber
\end{eqnarray}

Similarly, $R_{s's}(\tau)$ can be calculated. $\square$

If $s$ in Construction B  changes into $s'$  in Construction A, then $s'$ possesses the following properties of correlation:
\begin{theorem}\label{ss2} Let $\tau=\tau_1T+\tau_2,  0\leq \tau_2 \leq T-1$. The autocorrelation of $s'$ is given by
\begin{eqnarray*}
&&R_{s'}(\tau)
=\left\{
\begin{array}{lll}
R_{s}(\tau) & \mbox{ if } \tau_2=0,\\
R_{s}(\tau)-2d(a_{\tau_2})-2d(a_{T-\tau_2})&\mbox{ if } \tau_2\neq 0.
\end{array}
\right.
\end{eqnarray*}
The cross-correlation of sequences $s$ and $s'$ is given by
\begin{eqnarray*}
&&R_{ss'}(\tau)=\left\{
\begin{array}{lll}
TK-2K & \mbox{ if } \tau=0,\\
R_{s}(\tau)-2K&\mbox{ if } \tau_2=0, \tau\neq0,\\
R_{s}(\tau)-2d(a_{T-\tau_2})&\mbox{ otherwise };
\end{array}
\right. \\
&&R_{s's}(\tau)=\left\{
\begin{array}{lll}
TK-2K & \mbox{ if } \tau=0,\\
R_{s}(\tau)-2K&\mbox{ if } \tau_2=0, \tau\neq0, \\
R_{s}(\tau)-2d(a_{\tau_2})&\mbox{ otherwise }.
\end{array}
\right.
\end{eqnarray*}
\end{theorem}

As a consequent result of Theorem \ref{ss}, we have
\begin{corollary}\label{s2} For the sequences $s$ and $s'$,
\begin{eqnarray*}
R_{s's}(\tau)&=&R_{s s'}(\tau)\Longleftrightarrow d(a_{T-\tau_2})= d(a_{\tau_2}). \\
R_{s}(\tau)&=&R_{s'}(\tau)\Longleftrightarrow d(a_{T-\tau_2})=-d(a_{\tau_2}).
\end{eqnarray*}
\end{corollary}

On more special conditions, we have

\begin{corollary}\label{c0}Let $d(a_{T-\tau_2})+d(a_{\tau_2})=d_0$ be a constant.

(1) If $d_0=0$,  $s'$ possesses ideal autocorrelation if and only if  $s$  has ideal autocorrelation

(2) If $d_0$ is a nonzero constant, then $s'$  possesses 3-level autocorrelation if and only if  $s$  has ideal autocorrelation.
\end{corollary}

\begin{corollary} \label{th7}Let $d(a_{T-\tau_2})=d(a_{\tau_2})\neq K $ be a constant. $R_{ss'}(\tau)=R_{s's}(\tau)$ is 3-valued if and only if the sequence $s$  has ideal autocorrelation.
\end{corollary}

\begin{remark}Theorems \ref{ss}, \ref{ss2} and Corollaries \ref{s2}-\ref{th7} can induce that some binary sequences with good correlation can be obtained by changing $s$ into  $s'$  and its inverse process.
\end{remark}

Several known results will be introduced to verify  Corollaries \ref{s2}$-$\ref{th7}.

Let $p$ be an odd prime and $QR_p$ and $NQR_p$ denote the quadratic residue and nonquadratic residue of $p$.
A Legendre sequence $l(t)$ is defined as
$$l(t)=\left\{
\begin{array}{lll}
0\mbox{ or }1 &\mbox{if } t=0, \\
0 &\mbox{if } t\in QR_p, \\
1 &\mbox{otherwise}.
\end{array}
\right.
$$
$l(t)$ is called the first type Legendre sequence  if $l(0)=1$ otherwise the second type Legendre sequence (denoted by $l'(t)$).

\begin{lemma} \cite{tg} \label{R-legendre} Legendre sequences $l(t)$ and $l'(t)$ possess the following autocorrelation.
If $p\equiv3\bmod 4$, $l(t)$ and $l'(t)$ possess ideal autocorrelation, and if $p\equiv1\bmod 4$,
\begin{eqnarray*}R_l(\tau)=\left\{
\begin{array}{lll}
p    &\mbox{if } \tau=0, \\
1   &\mbox{if }  \tau\in QR_p, \\
-3  &\mbox{if }  \tau\in NQR_p.
\end{array}
\right.
R_{l'}(\tau)=\left\{
\begin{array}{lll}
p    &\mbox{if } \tau=0, \\
-3   &\mbox{if } \tau\in QR_p, \\
1   &\mbox{if } \tau\in NQR_p,
\end{array}
\right.
\end{eqnarray*}
\end{lemma}
and each type of Legendre sequences satisfies
\begin{eqnarray}\label{tpt}
s(t)-s(p-t)=0 &\mbox{ if } p\equiv1\bmod 4, \label{tpt1}\\
s(t)+s(p-t)=1 &\mbox{ if } p\equiv3\bmod 4,\label{tpt3}
\end{eqnarray}
where $t=1,2,\ldots,p-1$.

In \cite{Tingyao Xiong}, sequences satisfying Equations (\ref{tpt1}) and (\ref{tpt3}) are called symmetric and  antisymmetric respectively, and some new sequences with these properties are introduced. Obviously, these sequences can confirm the equivalences  in Theorem \ref{s2} respectively. Combining Equation $(\ref{tpt3})$ with (1) of Theorem \ref{c0} can  explain that these two types both possess ideal autocorrelation when $p\equiv3\bmod4$ \cite[Property 2]{tg}. Combining Equation $(\ref{tpt1})$, Lemma \ref{R-legendre} with  Theorem \ref{ss2} can  explain  the cross-correlations $R_{l'l}$ and $R_{ll'}$ are equal and 2-valued when $p\equiv1\bmod4$ \cite[Property 3]{tg}.

For the twin-prime sequence
$t=I(0_p, L^{e_1}(a_1)+b(1), \ldots L^{e_{p+1}}(a_{p+1})+b(p+1))$,
where $e_i=i(p+2)^{-1}\bmod p$, $p$ and $p+2$ are two primes, $b(i)=1$ if $i\in QR_{p+2}$ otherwise $b(i)=0$, and $a_i=l'$ if $i\in QR_{p+2}$ otherwise $a_i=l, i=1, 2, \ldots, p+1$. If $p\equiv1\bmod 4$, then $p+2\equiv3\bmod 4$, by Equation (\ref{tpt3}),
 $b(i)+b(p+2-i)=1$,
thus $b(i)=1$ if and only if $a_i=l'$ if and only if  $a_{p+2-i}=l$, and
\begin{eqnarray}\label{dlb}
 d(L^{e_i}(a_i)+b(i))
 =d(L^{e_{p+2-i}}(a_{p+2-i})+b(p+2-i))
 =1.
 \end{eqnarray}

If $p\equiv3\bmod 4$, then $p+2\equiv1\bmod 4$, and by Equation (\ref{tpt1}), $b(i)=b(p+2-i)=1$ if and only if $a_i=a_{p+2-i}=l'$. Thus Equation (\ref{dlb}) is also right.

The above Equation (\ref{dlb}) and Theorem \ref{th7} can  explain  the modified type
$
t'=I(1_p, L^{e_1}(a_1)+b(1), \ldots, L^{e_{p+1}}(a_{p+1})+b(p+1))
$
possesses 3-level autocorrelation and the equal 3-level cross-correlations $R_{t't}$ and $R_{tt'}$, which are the results of Property 5 in \cite{tg}.

It is well known that any binary sequence with ideal autocorrelation possesses balanced property, from  Theorem \ref{th7}, if $s$ is an classical interleaved sequence in construction A \cite{g}, then $s'$ possesses 3-level autocorrelation. Property 1 in \cite{tg} can be induced by this result.
Moreover, autocorrelation functions of all three generalized sequences $s'$s in \cite{tg} can be obtained by the above Theorem \ref{ss}

\section{Construction of New Sequences with Optimal Autocorrelation }

In \cite{tg}, a new interleaved sequence was defined as
$$u=I(s', L^{\frac{1}{4}+\eta}(s')+1,
L^{\frac{1}{2}}(s)+1, L^{\frac{3}{4}+\eta}(s)+1),
$$
where $s$ and $s'$ are interleaved binary sequences in  Constructions A and B respectively. Since the construction of $u$ is determined by the sequence $s$,  this section considers the relationship between their autocorrelation functions.

\begin{theorem}\label{ge}Let $\mu=4\mu_1+\mu_2, \mu_2=0, 1, 2, 3$.

(1) If $d(a_{x})=c_{1}, x=0, 1, \ldots, T-1$, then the  autocorrelation function of the sequence $u$ is given by
\begin{eqnarray*}
R_{u}(\mu)=
\left\{
\begin{array}{llllllll}
4KT & \mbox{ if } \mu=0, \\
4R_{s}(\mu_1)&\mbox{ if } \mu_2= 0, \tau_2=0, \mu\neq 0, \\
4R_{s}(\mu_1)+8c_1&\mbox{ if } \mu_2=0, \tau_2 \neq0, \\
0&\mbox{ if } \mu_2=1, \tau_1^{+}=0, \\
-4c_{1}&\mbox{ if } \mu_2=1, \tau_1^{+}\neq0, \\
0&\mbox{ if } \mu_2=2, \\
0&\mbox{ if } \mu_2=3, \tau_2^{-}=0, \\
-4c_{1}&\mbox{ if } \mu_2=3, \tau_2^{-}\neq0.
\end{array}
\right.
\end{eqnarray*}

(2) If $d(a_{x})+d(a_{T-x})=0, x=1, \ldots, T-1$, then the  autocorrelation function of the sequence $u$ is given by
\begin{eqnarray*}
R_{u}(\mu)
=\left\{
\begin{array}{llllllll}
4KT & \mbox{ if } \mu=0,\\
4R_{s}(\mu_1)&\mbox{ if } \mu_2= 0, \mu\neq 0. \\
0&\mbox{ if } \mu_2=1, \tau_{1}^{-}=0, \\
4d(a_{\tau_{2}^{-}})&\mbox{ if } \mu_2=1, \tau_{1}^{-}\neq0, \\
0&\mbox{ if }  \mu_2=2, \\
0&\mbox{ if } \mu_2=3, \tau_{2}^{+}=0, \\
-4d(a_{\tau_{2}^{-}})&\mbox{ if } \mu_2=3, \tau_{2}^{+}\neq0.
\end{array}
\right.
\end{eqnarray*}
\end{theorem}
\noindent\textbf{Proof. }
By Lemma \ref{0} and due to four different values of $\mu_2$,  the autocorrelation of the sequence $u$ can be given by the following:

Case 1: If $\mu_2=0$, then $R_u(\mu)=2R_{s'}(\mu_1)+2R_{s}(\mu_1)$.

Let $\mu_1=\tau_1T+\tau_2, 0\leq \tau_2\leq T-1$.
Then, by Theorems \ref{ss} and \ref{ss2}, we have\\
  (1)  if $\tau_2=0$, $R_u(\mu)=4R_{s}(\mu_1), $
  (2)  if $\tau_2\neq0$,
$R_u(\mu)=4R_{s}(\mu_1)+4d(a_{\tau_2})+4d(a_{T-\tau_2}). $

Case 2: If $\mu_2=1$, $R_u(\mu)
=R_{s}(\frac{1}{4}+\eta+\mu_1)-R_{s'}(\frac{1}{4}+\eta+\mu_1)
+R_{s's}(\frac{1}{4}-\eta+\mu_1)-R_{ss'}(\frac{1}{4}-\eta+\mu_1).
$

Let $
\frac{1}{4}+\eta+\mu_1\equiv\tau_{1}^{+}\bmod T,
\frac{1}{4}-\eta+\mu_1\equiv\tau_{1}^{-}\bmod T,
$
where $0\leq \tau_{1}^{+}, \tau_{1}^{-} \leq T-1$.
Then, by Theorems \ref{ss} and \ref{ss2}, we have
$$
R_{u}(\mu)=\left\{
\begin{array}{lll}
0 & \mbox{ if } \tau_{1}^{+}=\tau_{1}^{-}=0,\\
2d(a_{\tau_{1}^{-}})-2d(a_{T-\tau_{1}^{-}}) &\mbox{ if } \tau_{1}^{+}=0 \mbox{ and } \tau_{1}^{-}\neq0, \\
-2d(a_{\tau_{1}^{+}})-2d(a_{T-\tau_{1}^{+}})&\mbox{ if } \tau_{1}^{+}\neq0  \mbox{ and } \tau_{1}^{-}=0,\\
2d(a_{\tau_{1}^{-}})-2d(a_{T-\tau_{1}^{-}}) &\mbox{ if } \tau_{1}^{+}\neq0 \mbox{ and } \tau_{1}^{-}\neq0.\\
-2d(a_{\tau_{1}^{+}})-2d(a_{T-\tau_{1}^{+}})&
\end{array}
\right.
$$

Case 3: If $\mu_2=2$, then  \begin{eqnarray*}
R_u(\mu)
&=&-R_{s's}(\frac{1}{2}+\mu_1)+R_{s's}(\frac{1}{2}+\mu_1)
-R_{ss'}(-\frac{1}{2}+\mu_1)+R_{ss'}(-\frac{1}{2}+\mu_1)\\
&=&0.
\end{eqnarray*}

Case 4: If $\mu_2=3$, then
\begin{eqnarray*}
R_u(\mu)&=&R_{s}(\frac{3}{4}-\eta+\mu_1)-R_{s'}(\frac{3}{4}-\eta+\mu_1)\\
&&+R_{ss'}(\frac{3}{4}+\eta+\mu_1)-R_{s's}(\frac{3}{4}+\eta+\mu_1),
\end{eqnarray*}

Let $
\frac{3}{4}+\eta+\mu_1\equiv\tau_{2}^{+}\bmod T,
\frac{3}{4}-\eta+\mu_1\equiv\tau_{2}^{-}\bmod T,
$
where $0\leq \tau_{2}^{+}, \tau_{2}^{-} \leq T-1$.
By Theorems \ref{ss} and \ref{ss2}, we have
$$
R_{u}(\mu)=\left\{
\begin{array}{lll}
0 & \mbox{ if } \tau_{2}^{+}=0  \mbox{ and  }\tau_{2}^{-}=0,\\
-2d(a_{\tau_{2}^{-}})-2d(a_{T-\tau_{2}^{-}}) &\mbox{ if } \tau_{2}^{+}=0 \mbox{ and } \tau_{2}^{-}\neq0, \\
2d(a_{T-\tau_{2}^{+}})-2d(a_{\tau_{2}^{+}})&\mbox{ if } \tau_{2}^{+}\neq0  \mbox{ and } \tau_{2}^{-}=0,\\
2d(a_{T-\tau_{2}^{+}})-2d(a_{\tau_{2}^{+}}) &\mbox{ if } \tau_{2}^{+}\neq 0 \mbox{ and } \tau_{2}^{-}\neq0.\\
-2d(a_{\tau_{2}^{-}})-2d(a_{T-\tau_{2}^{-}})&
\end{array}
\right.
$$

Then, by Lemma \ref{0} and Theorem \ref{ss}, the proof can be completed. $\square$

As a direct corollary of Theorem \ref{ge}, we consider the following  case.
\begin{theorem}\label{3l} The sequence $u$ possesses optimal autocorrelation if and only if it satisfies either of the following conditions:

Condition 1: the sequence $s$ has ideal autocorrelation and $d(a_{x})=1$.

In this case, the autocorrelation function of the sequence $u$ is given by
\begin{eqnarray*}
R_{u}(\mu)
=\left\{
\begin{array}{llllllll}
4KT & \mbox{ if } \tau=0,\\
-4&\mbox{ if } \mu_2= 0, \tau_2=0, \mu\neq 0, \\
4&\mbox{ if } \mu_2=0, \tau_2\neq0,\\
0&\mbox{ if } \mu_2=1, \mu_1\equiv-\frac{1}{4}-\eta \bmod T, \\
-4&\mbox{ if } \mu_2=1, \mu_1\not\equiv-\frac{1}{4}-\eta \bmod T, \\
0&\mbox{ if }  \mu_2=2, \\
0&\mbox{ if } \mu_2=3, \mu_1\equiv-\frac{3}{4}+\eta \bmod T, \\
-4&\mbox{ if } \mu_2=3, \mu_1\not\equiv -\frac{3}{4}+\eta \bmod T.
\end{array}
\right.
\end{eqnarray*}
Condition 2:  $s$ has ideal autocorrelation and $d(a_{x})=-d(a_{T-x})\in\{1,-1\}$.

In this case, the  autocorrelation function of the sequence $u$ is given by
\begin{eqnarray*}
R_{u}(\mu)=\left\{
\begin{array}{llllllll}
4KT & \mbox{ if } \mu=0, \\
-4&\mbox{ if } \mu_2= 0, \mu\neq 0, \\
0&\mbox{ if } \mu_2=1, \tau_1\equiv-\frac{1}{4}+\eta \bmod T, \\
\pm4&\mbox{ if } \mu_2=1, \mu_1 \not\equiv -\frac{1}{4}+\eta \bmod T, \\
0&\mbox{ if }  \mu_2=2, \\
0&\mbox{ if } \mu_2=3, \mu_1\equiv-\frac{3}{4}-\eta \bmod T, \\
\mp4&\mbox{ if } \mu_2=3, \mu_1 \not\equiv -\frac{3}{4}-\eta \bmod T.
\end{array}
\right.
\end{eqnarray*}
\end{theorem}
\vspace{5mm}
Actually, all three constructions of sequences with optimal autocorrelation in \cite{tg} can be included in the above Theorem \ref{3l}. More specifically, autocorrelation of Constructions A and B in \cite{tg} can be explained by the equivalence about Condition 1 of the above Theorem \ref{3l}, and the equivalence about Condition 2 can explain autocorrelation of main parts of Construction C in \cite{tg} directly. Moreover, based on our Theorem \ref{ge}, many binary sequences with low autocorrelation can be constructed by searching more binary sequences with low autocorrelation in Constructions A and B.
\section{Acknowledgement}
This paper was completed while the first author was a visiting scholar at the Department of ECE of University of Waterloo. We would like to express our gratitude to Professor G. Gong. for the supports.

\end{document}